\documentclass{appolb}
\usepackage{epsfig}
\begin{document}
\title{High$-p_{T}$ processes measured with ALICE at the LHC
\thanks{Presented at Strangeness in Quark Matter 2011}
}
\author{Jacek Otwinowski for the ALICE Collaboration
\address{GSI Helmholtz Centre for Heavy Ion Research GmbH, Planckstrasse 1, 64291~Darmstadt, Germany}
}
\maketitle
\begin{abstract}
From studies of single-particle spectra, particle correlations and jet production, in heavy-ion collisions we can obtain information about the density and the dynamic properties of the Quark Gluon Plasma (QGP). The observed suppression of high$-p_{T}$ particle production ($R_{AA}$) and away-side jets ($I_{AA}$) is generally attributed to energy loss of partons as they propagate through the plasma. We present the results obtained from the analysis of Pb$-$Pb collisions at $\sqrt{s_{NN}} = 2.76$ TeV recorded by ALICE in November 2010. The nuclear modification factors $R_{AA}$ and $I_{AA}$, and the status of full jet reconstruction in Pb$-$Pb is presented. Comparison with the RHIC measurements at lower collision energy and with theory models is shown.
\end{abstract}
\PACS{25.75.-q}

\section{Introduction}

This paper reports on measurements of single-particle inclusive spectra and two-particle azimuthal correlations for hadrons  as a function of transverse momentum ($p_{T}$) and event centrality in Pb$-$Pb collisions at $\sqrt{s_{NN}}=2.76$~TeV recorded  by ALICE  {\cite{ALICE_DET}} in November 2010. 

The measurement is motivated by results \cite{STAR_RAA_IAA, BRAHMS_RAA,PHOBOS_RAA,STAR_RAA,PHENIX_RAA} from the Relativistic Heavy Ion Collider (RHIC), which showed that hadron production at large transverse momentum in central Au$-$Au collisions at $\sqrt{s_{NN}}=200$~GeV is suppressed by a factor $4-5$ compared to expectations from an independent superposition of nucleon-nucleon collisions. This observation is typically expressed in terms of the nuclear modification factor which is defined as the ratio of the charged particle yield in Pb$-$Pb to that in pp, scaled by the number of binary nucleon-nucleon collisions $\langle N_{coll} \rangle$

\begin{equation}\label{EQUATION_RAA}
R_{AA}(p_{T}) = \frac{(1/N^{AA}_{evt})d^{2}N^{AA}_{ch}/d\eta dp_{T}}{\langle N_{coll} \rangle (1/N^{pp}_{evt})d^{2}N^{pp}_{ch}/d\eta dp_{T}}
\end{equation}
In absence of nuclear modifications $R_{AA}$ is equal to unity at high $p_{T}$. 
 
At RHIC it was also measured \cite{STAR_IAA} that the back-to-back di-hadron correlations are considerably reduced in the most central Au$-$Au collisions as compared to pp, indicating a substantial interaction as the hard-scattered partons or their fragmentation products traverse the medium. It is quantified by comparison of hadron yields measured in Pb-Pb ($Y_{AA}$) and pp ($Y_{pp}$) collisions

\begin{equation}\label{EQUATION_IAA}
I_{AA}(p_{T},\Delta\phi) = \frac{Y_{AA}(p_{T},\Delta\phi)}{Y_{pp}(p_{T},\Delta\phi)}
\end{equation}
 
The $Y_{AA}$ and $Y_{pp}$ are extracted from the background subtracted per-trigger particle yield ($D(\Delta\phi)=dN_{assoc}/N_{trig}d(\Delta\phi$)) where azimuthal correlation is built between a high$-p_{T}$ triggered leading hadron ($p_{T,trig}>p_{T,thresh}$) and all associated particles ($p_{T,assoc}<p_{T,trig}$) in one event.  In absence of nuclear modifications the $I_{AA}$ is equal to unity by construction.

\begin{figure}[t]
\begin{center}$
\begin{array}{cc}
\includegraphics[width=2.4in]{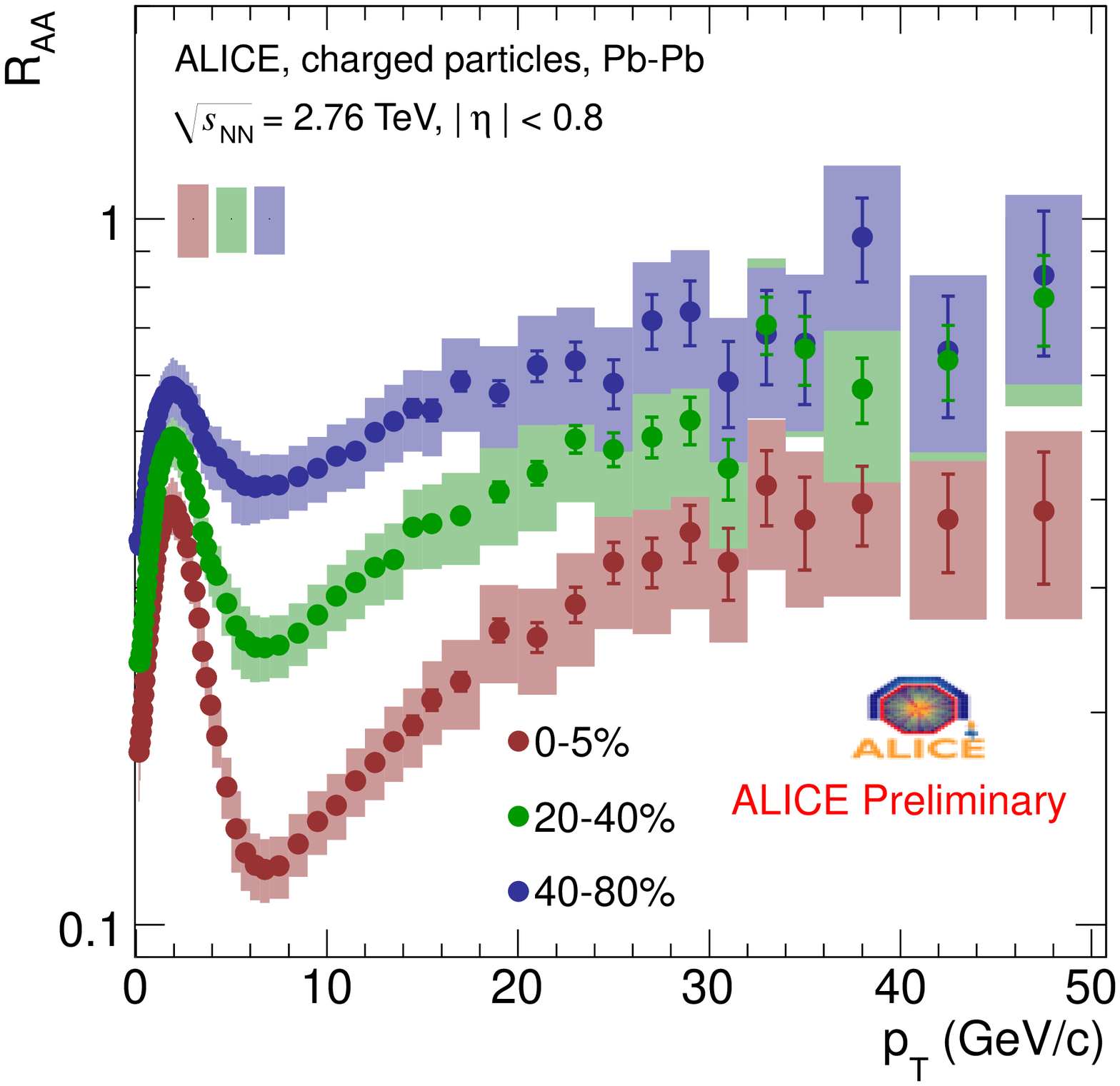} &
\includegraphics[width=2.4in]{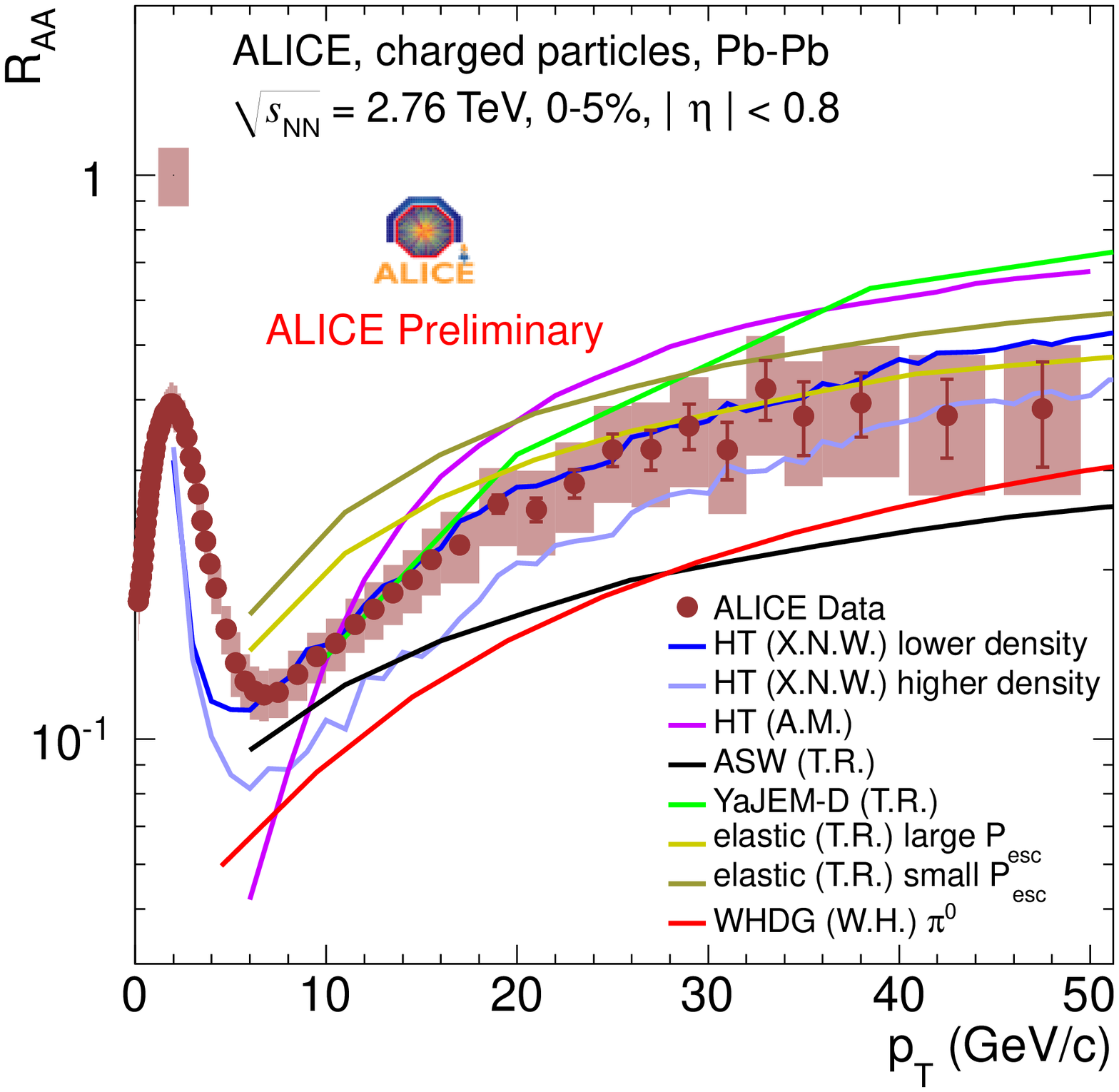}
\end{array}$
\end{center}
\caption{\label{RAASPECTRA} {\bf Left:}   $R_{AA}$ of charged particles measured with ALICE in central Pb$-$Pb collisions in three centrality intervals. {\bf Right:} $R_{AA}$ of charged particles measured with ALICE in central Pb$-$Pb collisions ($0-5\%$) in comparison to model calculations. The error bars at $R_{AA}$=1 denote contributions from normalization uncertainties.}
\end{figure}

\section{Results}

\subsection{$R_{AA}$ charged particles}

\begin{figure}[t]
\begin{center}$
\begin{array}{cc}
\includegraphics[width=6.2cm,height=6.cm]{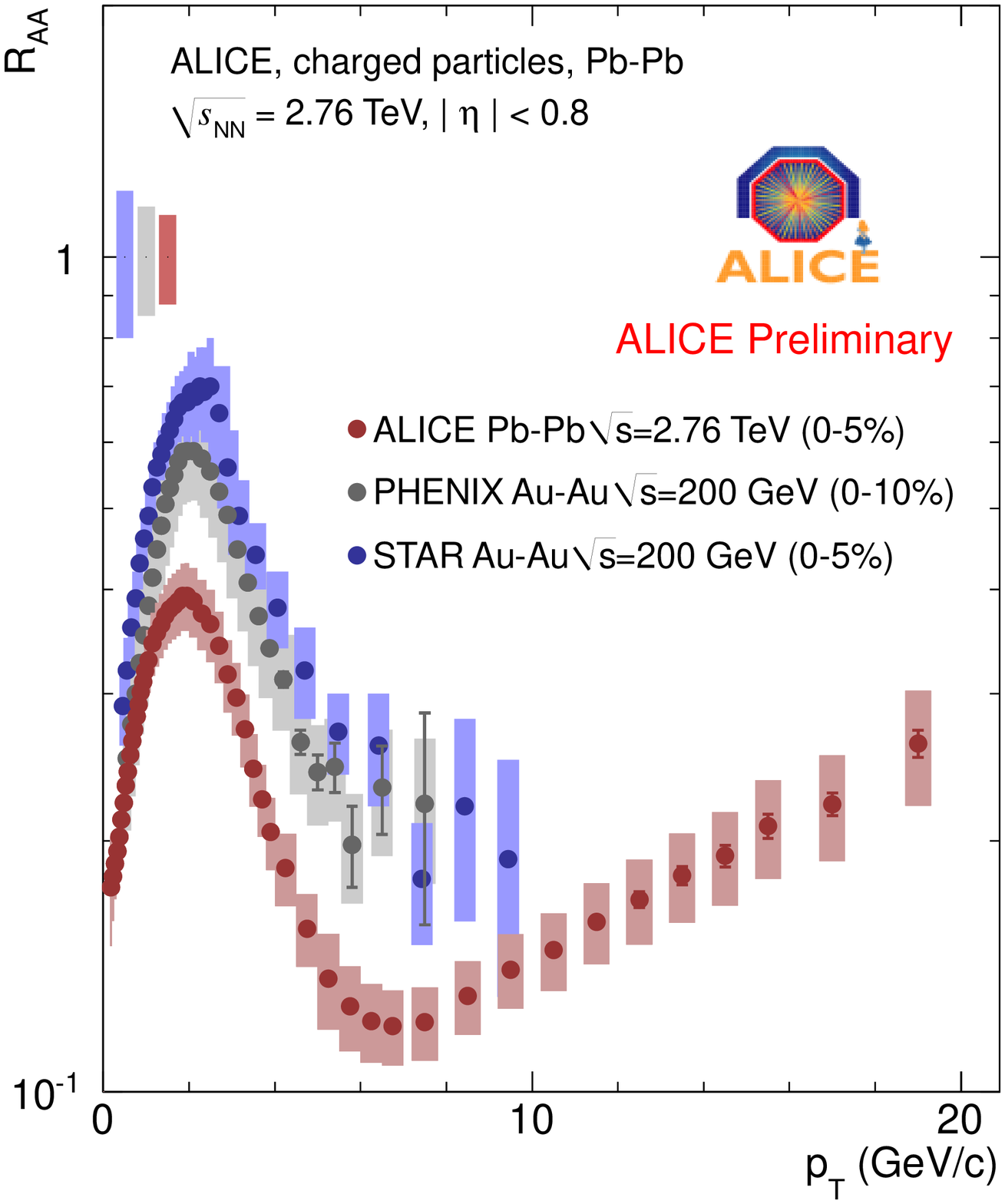} &
\includegraphics[width=6.0cm,height=6.0cm]{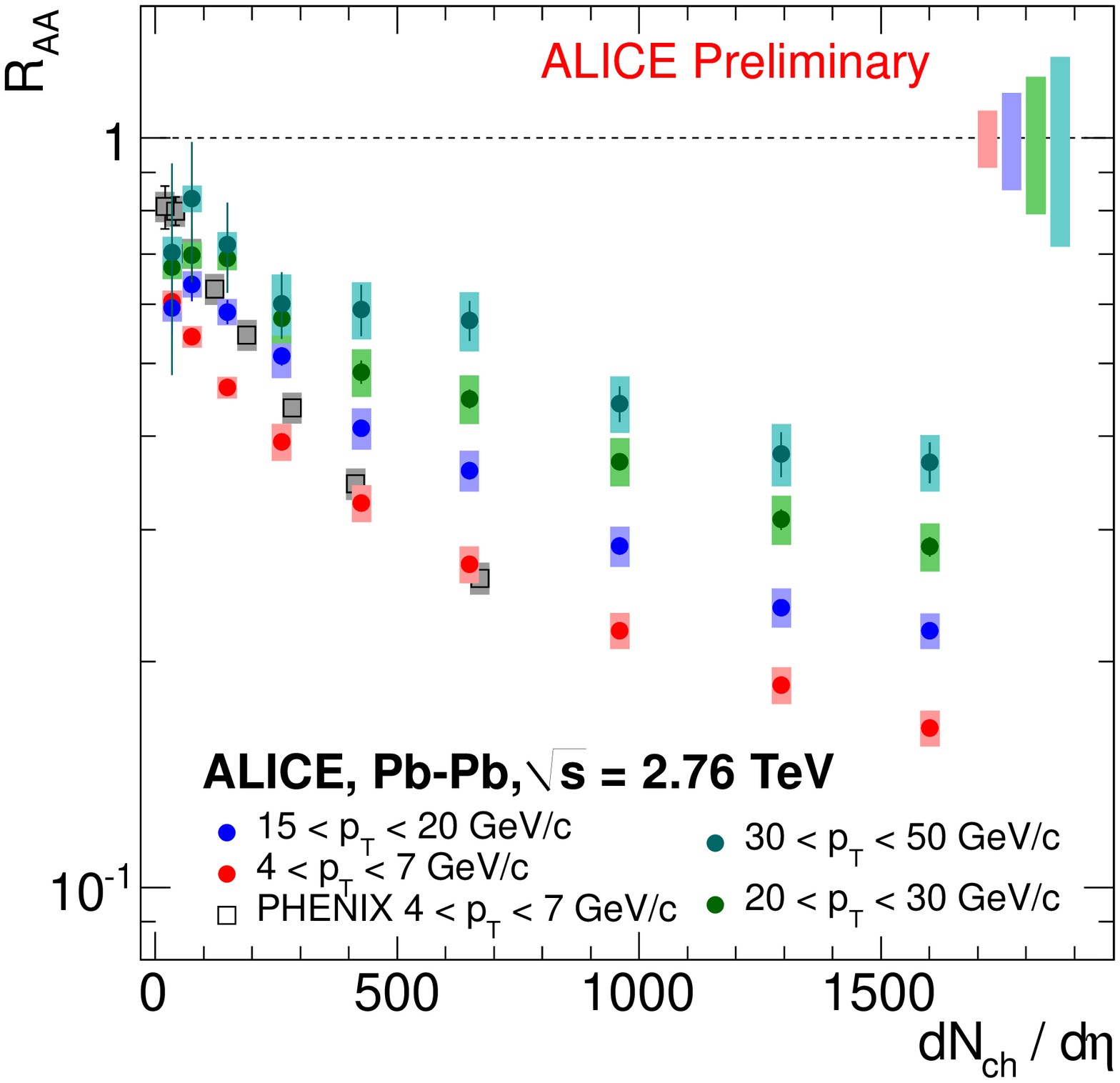}
\end{array}$
\end{center}

\caption{\label{RAARHIC} {\bf Left:}  $R_{AA}$ of primary charged particles measured with ALICE in central Pb$-$Pb collisions ($0-5\%$) in comparison to RHIC measurements. The error bars at $R_{AA}$=1 denote the contributions from normalization uncertainties.  {\bf Right:} $R_{AA}$ of  charged particles showed as a function of particle density ($dN_{ch}/d\eta$) in four $p_{T}$ intervals. Comparison to PHENIX measurements is also shown. The error bars at $R_{AA}$=1 denote $p_{T}$-dependent systematic uncertainties on pp reference spectrum. The normalization uncertainties on pp spectrum (3.5\% independent of $p_{T}$) are not plotted.}
\end{figure}

The nuclear modification factors out to  $p_{T}=50$~GeV/c are shown in Fig. {\ref{RAASPECTRA}} (left panel) for different centrality intervals. At all centralities, a pronounced minimum at
about $p_{T}=6-7$~GeV/c is observed. For $p_{T}>7$~GeV/c there is a significant rise in the nuclear modification factor until $p_{T}=30$~GeV/c.  This emphasizes the strong relation between the medium density and partonic energy loss. 

Fig. {\ref{RAASPECTRA}} (right panel) shows a comparison of $R_{AA}$ in central Pb$-$Pb collisions  ($0-5\%$) at  $\sqrt{s_{NN}}$=2.76~TeV to calculations from energy loss models {\cite{TH_1,TH_2, TH_3,TH_4}}. All model calculations have been constrained to match $R_{AA}$ results from RHIC. The qualitative features of our data are described by all models, including the strong
rise of $R_{AA}$ below $p_{T}=30$~GeV/c and the flattening off at higher $p_{T}$. A more quantitative comparison of model calculations to the present data will help to put tighter constraints on the underlying energy loss mechanisms. 

In Fig. \ref{RAARHIC} (left panel) $R_{AA}$ measured with ALICE is compared to measurements at lower collision energy ($\sqrt{s_{NN}} = 200$~GeV) by the PHENIX and STAR experiments {\cite{PHENIX_RAA_2, STAR_RAA_2}} at RHIC. At $p_{T}=1$~GeV/c the measured value of $R_{AA}$ is similar to those of RHIC. The position and shape of the maximum at $p_{T}\approx2$ GeV/c and the subsequent decrease are similar at RHIC and LHC. At $p_{T} $= 6--7~GeV/c  $R_{AA}$ is smaller than at RHIC indicating that a very dense medium is formed in Pb$-$Pb collisions at the LHC.

The $R_{AA}$ measured as a function of charged particle density $dN_{ch}/d\eta$ (from \cite{ALICE_dNdeta}) in four $p_{T}$ intervals is shown in Fig. \ref{RAARHIC} (right panel). The strongest $R_{AA}$ dependence as a function of $dN_{ch}/d\eta$ is observed for low$-p_{T}$ particles ($4<p_{T}<7$~GeV/c), and becomes weaker with larger $p_{T}$. There is also shown the PHENIX $R_{AA}$ measurement in the lowest $p_{T}$ interval. It is comparable with $R_{AA}$ measured by ALICE for $dN_{ch}/d\eta>400$. For smaller $dN_{ch}/d\eta$, PHENIX and ALICE measurements differ what might be related to the collision geometry and have to be further studied.

More analysis details related to $R_{AA}$  can be found in \cite{ALICE_RAA, ALICE_RAA_JACEK}.

\subsection{$R_{AA}$ identified hadrons}

\begin{figure}[t]
\begin{center}$
\begin{array}{cc}
\includegraphics[width=2.45in]{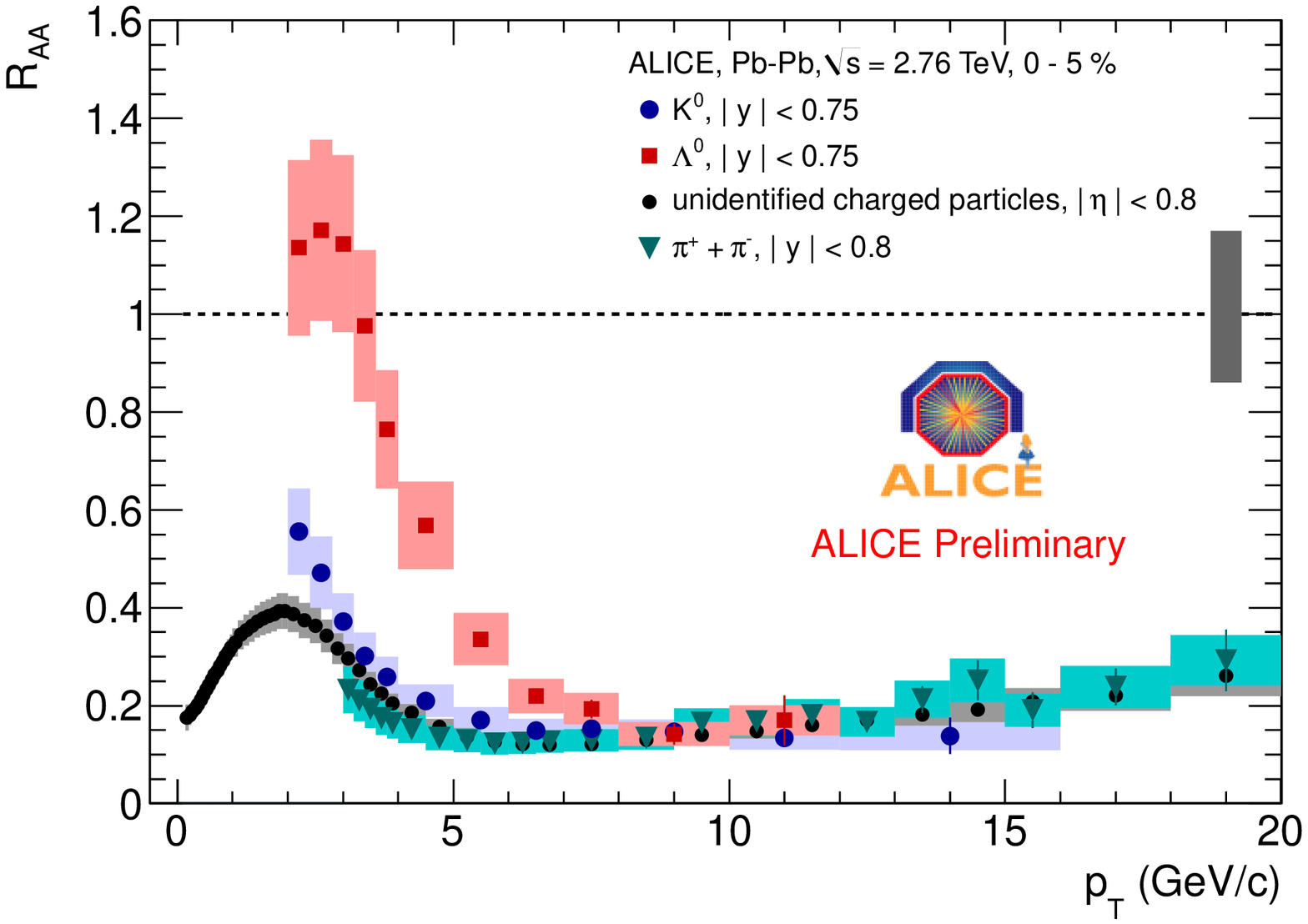} &
\includegraphics[width=2.45in]{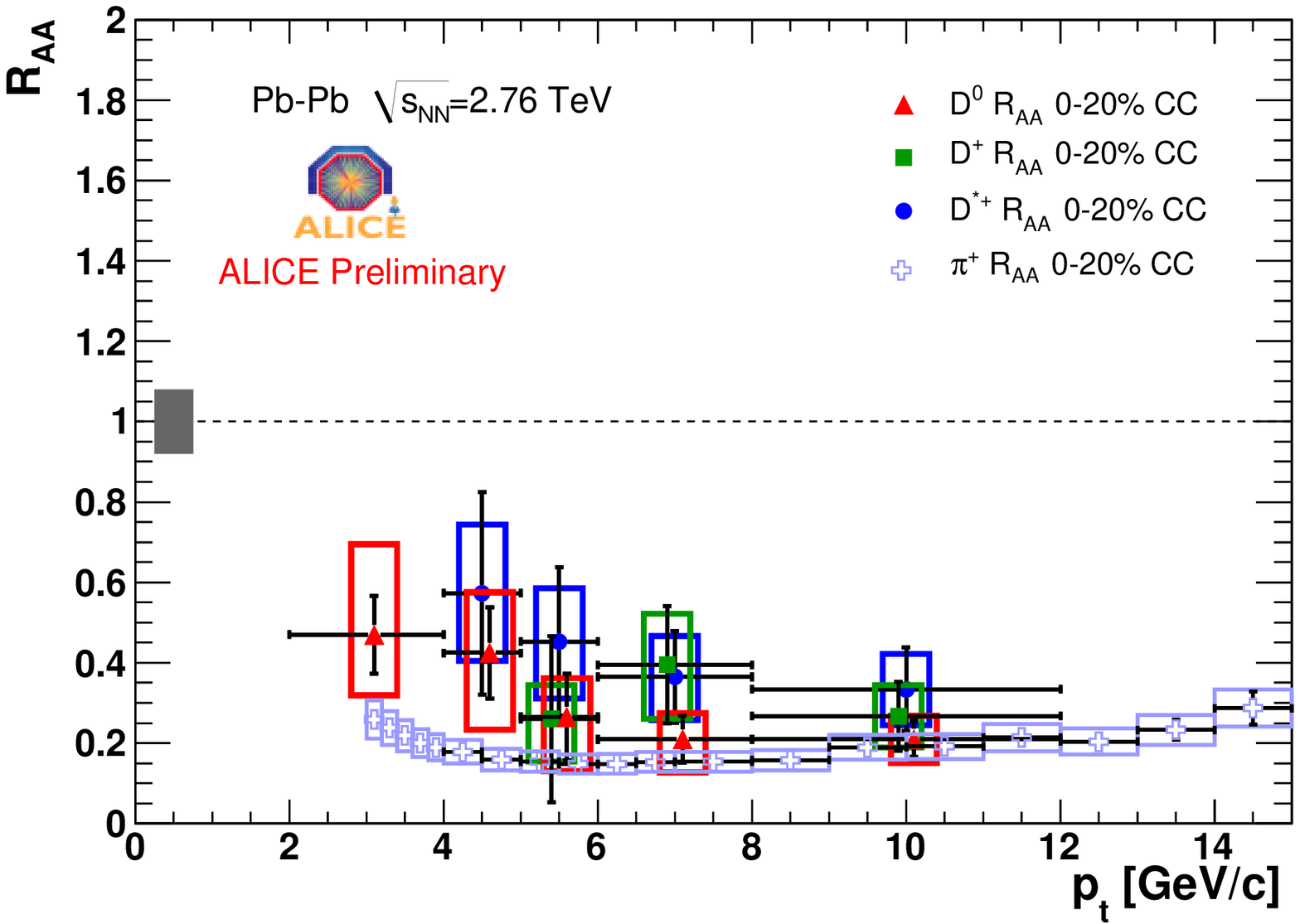}
\end{array}$
\end{center}
\caption{\label{RAAFLAVOR} {\bf Left:}   $R_{AA}$ of charged particles, $\pi$, $K^{0}_{s}$ and $\Lambda$ measured with ALICE in the central Pb-Pb collisions ($0-5\%$). {\bf Right:} $R_{AA}$ of charged pions and D mesons measured with ALICE in the central Pb$-$Pb collisions ($0-20\%$). The error bars at $R_{AA}$=1 denote contributions from normalization uncertainties.}
\end{figure}

The analysis of charged pions at high $p_{T}$ is based on statistical particle identification using the
specific energy loss dE/dx in the TPC \cite{PIONID}. In the region of the relativistic rise of the energy
loss ($p_{T} > 3$~GeV/c), the separation of pions from kaons and protons is nearly independent
of $p_{T}$ out to $p_{T} = 50$~GeV/c. The fraction of pions from all charged particles is determined
in bins of $p_{T}$ by fitting the dE/dx distribution with four Gaussians for p, K, $\pi$ and e.

The reconstruction of weak decays $K^{0}_{s}$ $\rightarrow$  $\pi^{+}$ + $\pi^{-}$ and  $\Lambda$  $\rightarrow$ $p + \pi$  \cite{STRANGEHADRONS}
allows us to study different suppression patterns for baryons and mesons, which may give a handle on how to separate quark and gluon energy losses. In Fig. \ref{RAAFLAVOR} (left panel) the $R_{AA}$ for $K^{0}_{s}$ and $\Lambda$ are shown in central Pb$-$Pb collisions in comparison to inclusive charged particles and pions out to $p_{T} = 16$~GeV/c. For $p_{T} > 6$~GeV/c, a significant suppression for $K^{0}_{s}$ and $\Lambda$ is seen which is similar to that of inclusive charged particles and pions. At lower $p_{T}$, the $R_{AA}$ of $\Lambda$ is significantly larger than that of $K^{0}_{s}$ which is in line with the observation of a strong and centrality$-$dependent enhancement of $\Lambda/K^{0}_{s}$ \cite{STRANGEHADRONS_2}.
 
The measurement of heavy-favor production at high-$p_{T}$ provides unique observables of jet
quenching. The suggestion that massive quarks experience reduced energy loss due to the
suppression of forward radiation ("dead cone effect") has not been borne out by RHIC
measurements. The first results from the LHC indicate (Fig. \ref{RAAFLAVOR}, right panel) that the suppression of D mesons might be smaller than for pions at $p_{T}<5$ GeV/c (more analysis details can be found in  {\cite{ALICE_RAA_ANDREA}}). Higher statistics of Pb$-$Pb data and comparison data in p$-$Pb collisions should allow to study this region with more precision and disentangle the initial-state nuclear effects, which could be different for light and heavy flavor. 

 \subsection{$I_{AA}$ charged particles}

\begin{figure}[t]
\begin{center}
\includegraphics[width=5.2in]{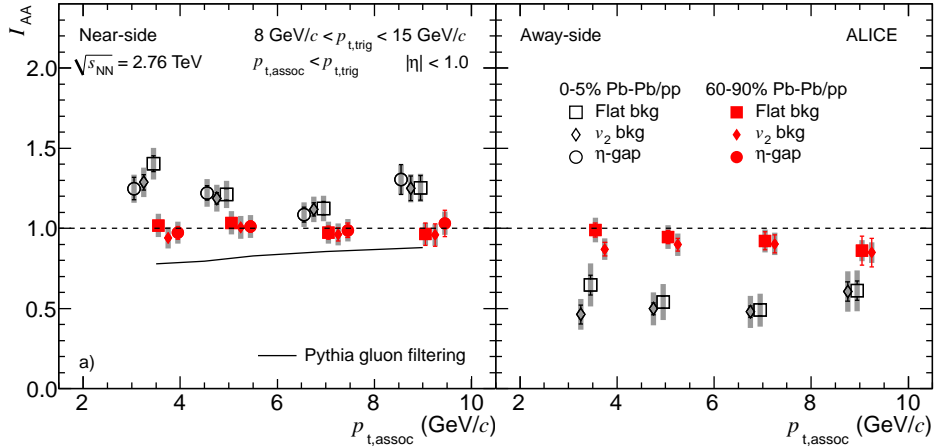} 
\end{center}
\caption{\label{IAASPECTRA}  $I_{AA}$ of charged particles measured with ALICE in Pb$-$Pb collisions in two centrality intervals ($0-5\%$,  $60-90\%$). For better visibility, the data points are slightly displaced on the $p_{T,assoc}$--axis. The shaded bands denote systematic uncertainties. The line (near--side) shows a PYTHIA 8 calculations illustrating the effect of gluon filtering in the medium (more details  in \cite{ALICE_IAA}).}
\end{figure}

The di--hadron back--to--back azimuthal correlations provide more differential information about  in--medium parton energy loss as compared to $R_{AA}$.   
It is quantified by measuring the ratio of  yields in Pb--Pb and pp collisions ($I_{AA}$) which are extracted from the near--side ($\Delta\phi\approx0$) and away--side ($\Delta\phi\approx\pi$) peaks.  

Fig. \ref{IAASPECTRA} shows  the $I_{AA}$ for central ($0-5\%$) and peripheral ($60-90\%$) collisions after background subtraction which is based on three different schemes: flat pedestal,  $v_{2}$ and $\eta$--gap (more details in \cite{ALICE_IAA}). The significant difference between $I_{AA}$ values is visible in the lowest $p_{T,assoc}$ bin what confirms a small bias due to the flow anisotropies in this
$p_{T}$ region. In central collisions, a strong yield suppression is observed on the away--side ($I_{AA}\approx0.6$) which is consistent with in--medium parton energy loss.  On the other hand, there is an unexpected yield enhancement ($I_{AA}\approx1.2$) on the near--side which has not been observed at lower collision energies \cite{STAR_IAA}.  In peripheral collisions, the yields are not modified and $I_{AA}$ is consistent with unity on both the near-- and away--side.

 \subsection{Full jet reconstruction in ALICE}

\begin{figure}[t]
\begin{center}$
\begin{array}{cc}
\includegraphics[width=2.55in]{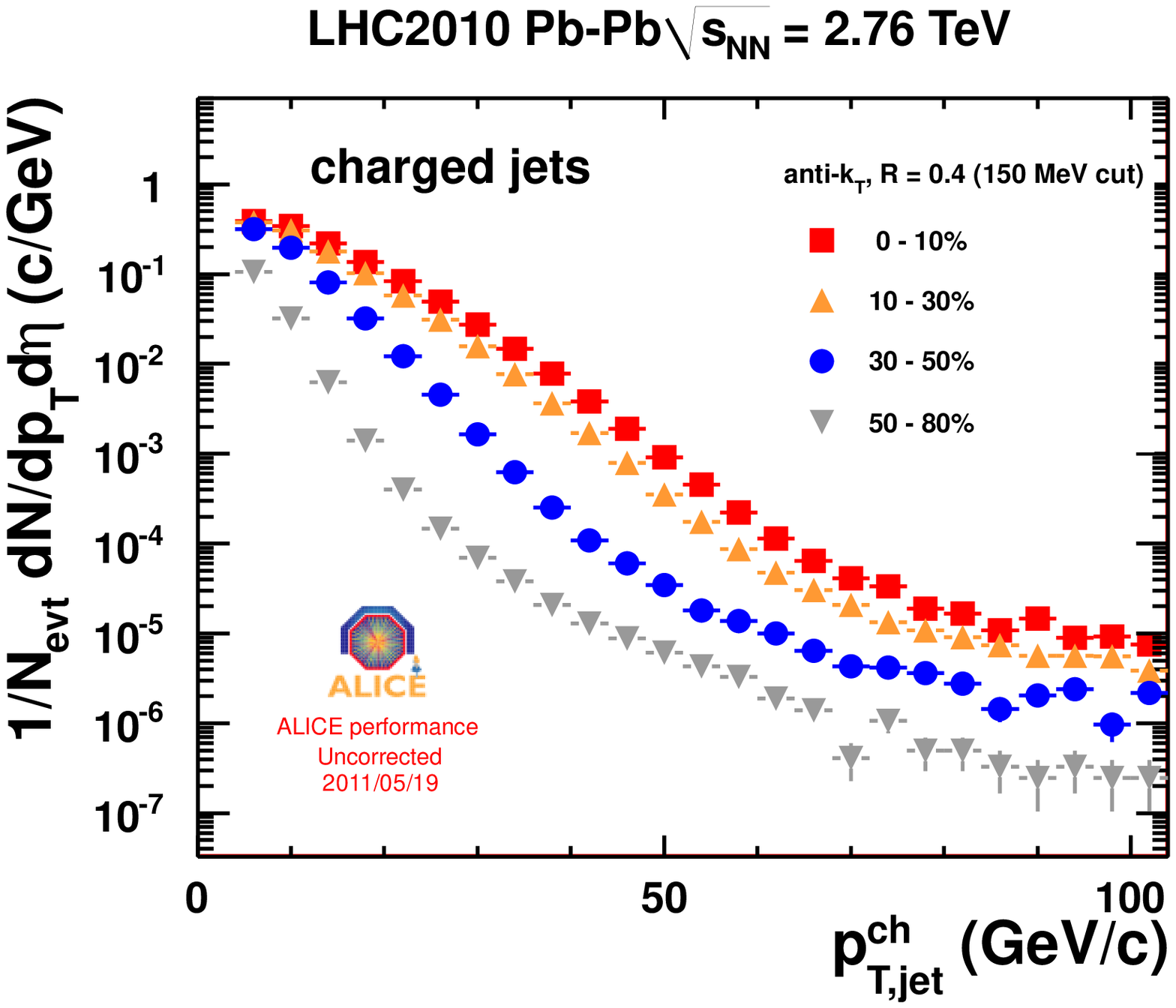} &
\includegraphics[width=2.55in]{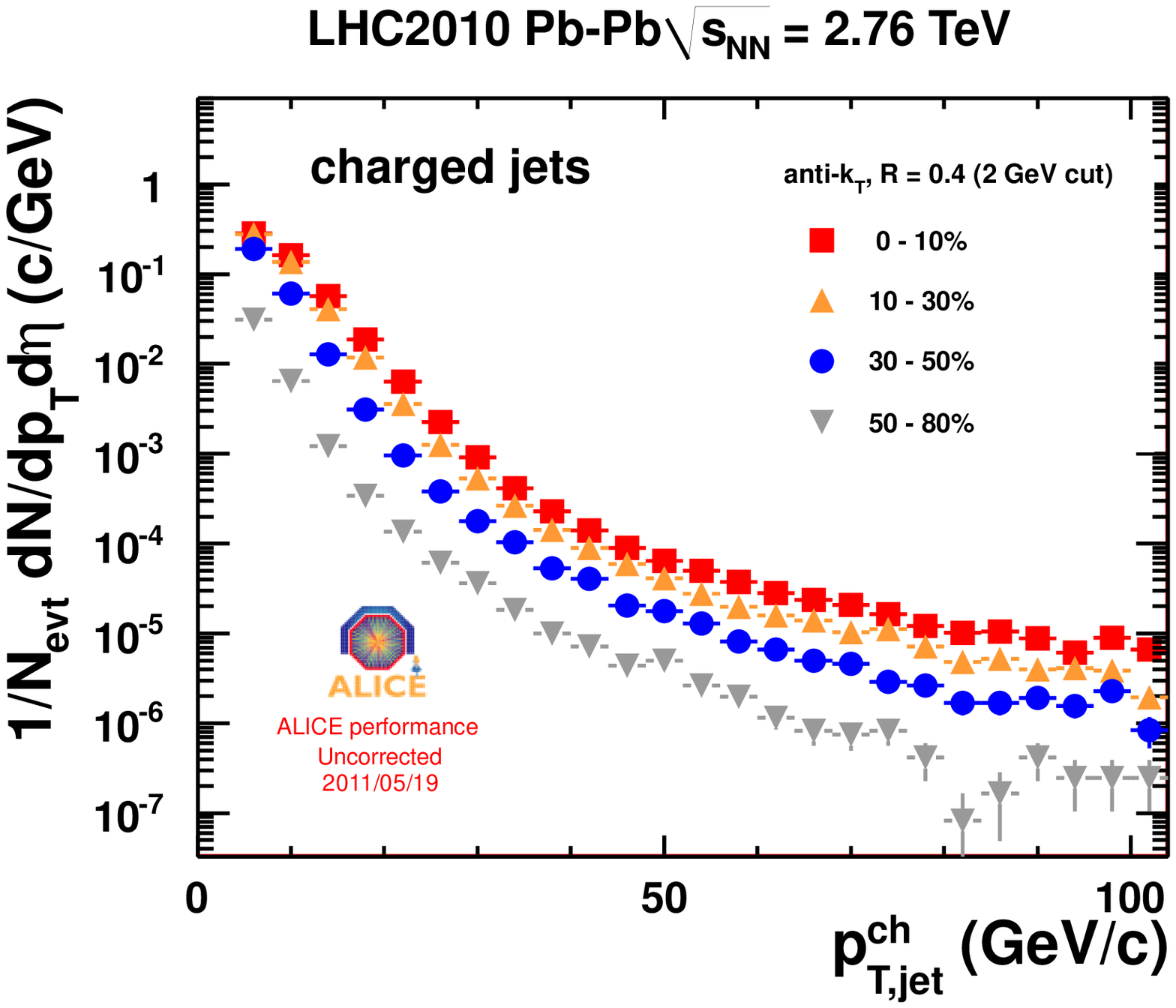}
\end{array}$
\end{center}
\caption{\label{JETSPECTRA} Reconstructed raw charged jet spectra in four centrality intervals using the anti$-k_{T} $ algorithm ($R=0.4$) after background subtraction (details in the text). Shown are spectra for two different track $p_{T}$ cut-off values:  $p_{T}>150$~MeV/c (left) and $p_{T}>2$~GeV/c (right), respectively.}
\end{figure}

The ALICE electromagnetic calorimeter (EMCal) \cite{ALICE_DET} was fully installed in January 2011. Thus jets from the first Pb--Pb collisions in ALICE (2010 run) are reconstructed based on charged particles only. The tracks are reconstructed using the Time Projection Chamber (TPC) and vertexing 
information from the Inner Tracking System (ITS). This ensures maximum azimuthal angle ($\phi$)
uniformity of reconstructed tracks with transverse momenta down to $p_{T}=150$~MeV/c. 

The full jets are reconstructed using the anti$-k_{T}$ algorithm \cite{ANTIKT} and are corrected for the background in each event using the jet area $A$ with $p^{ch}_{T,jet}  = p^{rec}_{T,jet} - \rho \cdot A$. Here, the background density $\rho$ is calculated on an event-by-event basis as the median for the $p_{T}/A$ ratio of reconstructed
$k_{T}-$clusters in $|\eta|< 0.5$ by using $k_{T}$ algorithm \cite{KT}. The resulting raw jet spectra are shown in Fig. \ref{JETSPECTRA} for two different track $p_{T}$ cut--off values.  The difference in the shape of the $p_{T}$ distributions in central collisions, clearly visible for the small track $p_{T}$ cut--off, is due to the background fluctuations which dominate over a wide $p_{T} $ range. The analysis of the jet background sources in Pb--Pb collisions is ongoing (more details in \cite{ALICE_JETS}).

\section{Summary}

The results indicate a strong suppression of charged particle production in Pb$-$Pb
collisions at $\sqrt{s}=2.76$~TeV and a characteristic centrality and $p_{T}$ dependence of the nuclear modification factors $R_{AA}$ and $I_{AA}$. The suppression observed in central Pb$-$Pb collisions at the LHC is
stronger than in central Au$-$Au collisions at RHIC. The comparison of ALICE data to model calculations indicates a large sensitivity of high-$p_{T}$ particle production to details of energy loss mechanisms.


\end{document}